**Title:** Systematic control of Raman scattering with topologically induced chirality of light

**Authors:** Xiao Liu, Zelin Ma, Aku Antikainen, Siddharth Ramachandran*

**Affiliation:**

Boston University, Boston, MA 02215, USA.

*Correspondence to: sidr@bu.edu

**Abstract:**

Stokes Raman scattering is known to be a particularly robust nonlinearity, occurring in virtually every material, with spectra defined by the material and strengths dependent on the material as well as light intensities. This ubiquity has made it an indispensable tool in spectroscopy, but also presents itself as a stubborn source of noise or parasitic emission in several applications. Here, we show that orbital angular momentum carrying light beams experiencing spin-orbit interactions can fundamentally alter the selection rules for Raman scattering. This enables tailoring its spectral shape (by over half the Raman shift in a given material) as well as strength (by ~ 100x) simply by controlling light's topological charge – a capability of utility across the multitude of applications where modulating Raman scattering is desired.

**Introduction:**

The physics and selection rules for Raman Stokes scattering have now been well known for almost a century – at a molecular level, it represents inelastic scattering between photons via energy transfer with lattice vibrations (phonons), and this effect can be quantitatively analyzed as a nonlinear optical scattering phenomenon with a medium possessing complex third order susceptibility. The growth of a Stokes wave can be written as [1]:

$$\frac{dI_s}{dz} = g_R(\Omega) I_p I_s \qquad (1)$$

where $z$ is the propagation distance, $I_p$ and $I_s$ are beam intensities of the pump and Stokes light, respectively, and $g_R(\Omega)$ is the frequency dependent Raman gain coefficient arising from the imaginary component of a material's nonlinear susceptibility. While this equation describes the behavior of Raman scattering at a cross-sectionally localized point in space, in general, the overall Raman strength for arbitrary beam shapes is also proportional to an intensity overlap integral $\eta$ over the transverse plane [2], given by:

$$\eta = \int I_p(r,\phi) \cdot I_s(r,\phi) r dr d\phi \qquad (2)$$

where $r$ and $\phi$ are radial and azimuthal coordinates. In addition, Raman gain is maximized for co-polarized light, hence in chiral media, a circularly polarized pump induces unequal amounts of Raman signal in the two opposite spins, yielding what is known as Raman optical activity [3]. Therefore, Raman Stokes scattering typically occurs for co-polarized light and its strength is proportional to an *intensity* overlap integral – i.e., it is impervious to the phase of either the pump or Stokes light. While Raman scattering dynamics for ultrashort pulses yield distinctive effects such as the soliton self-frequency shift and soliton self-mode conversion [4-6] where the effect also depends on the group velocity and group-velocity dispersion of light in the medium, the underlying rules – spectral response being governed by $g_R(\Omega)$, effect being *insensitive* to light's phase, and being maximized for co-polarized light – remain the same.

This is in contrast to well-known parametric nonlinearities such as second harmonic generation, four-wave-mixing [1] (FWM), or coherent anti-Stokes Raman scattering [7] (which is a special case of FWM), where phases of individual beams matter, and hence phase matching among the interacting photons determines the emission spectra [Fig. 1(a)]. The fundamental distinction is between elastic and inelastic scattering, Raman being of the latter variety, because a participating optical phonon mediates both energy and momentum conservation. Since the magnitude of momentum carried by lattice vibrations is orders of magnitude greater than that carried by photons, optical phonon wavevectors at almost any arbitrary orientations in relation to those of the incident and scattered photons are able to offer momentum conservation for the scattering process [Fig. 1(b)]. Hence the Raman scattering process automatically conserves momentum, and since momentum is intimately tied to the phase of a photon, Raman is typically known to be a self-phase-matched process. This imperviousness to phase is actually a key attribute for Raman Stokes scattering, since spectral shapes of absorption lines are not influenced by properties of the pump beam such as the angle of incidence of light, or the dispersive properties of matter being probed etc., hence yielding a highly reliable fingerprinting tool for spectroscopy [8]. But this also means that it is inevitably present in all systems and can be detrimental in applications such as quantum source generation [9] or power-scaling of fiber lasers [10], where it either acts as a noise source or reduces device efficiency. Thus, reported schemes for suppressing Raman scattering have involved 'brute-force' methods such as reducing beam intensity [11], inhibiting transmission at Stokes wavelengths [12] or cooling the material [13] to decrease the magnitude of $g_R(\Omega)$. Even so, none of these techniques can significantly alter the spectral shape $g_R(\Omega)$, defined solely by the material. Here, we show that light carrying orbital angular momentum (OAM) that experiences spin-orbit coupling, as readily occurring due to the confining potential of an optical fiber waveguide, can fundamentally alter Stokes Raman scattering by making it sensitive to the phases of the participating beams. This enables, for the first time to our knowledge, dispersive control of the strength as well as spectrum of Raman Stokes scattering simply by tailoring the topological charge of light.

Eigenmodes in a circularly symmetric waveguide such as an optical fiber [14] carry both orbital angular momentum, which is quantified by the topological charge $L$, and spin angular momentum $\hat{\sigma}^{\pm}$ (left/right-handed circular polarization). Spin-orbit interaction (SOI), originating from the light-matter interaction at dielectric interfaces, splits the effective refractive index ($\Delta n_{eff}$) between $\hat{\sigma}^{\pm}$ states of same $L$, inducing optical activity [15] (OA). The ring-core fiber we use in our experiments [Fig. 1(c)] amplifies this SOI, enabling stable propagation of each $\{L, \hat{\sigma}^{\pm}\}$ state. In addition, since this splitting is geometrodynamic in nature, optically active superpositions of these states, such as linearly polarized OAM modes, are also stable during propagation [15]. As a result, the polarization orientation angle of linearly polarized light carrying OAM at the input of the fiber systematically rotates as it propagates, with the beat length $Z_{beat} = \lambda/\Delta n_{eff}$. Schematic plots of this polarization evolution are shown in Figs. 1(d) and 1(e), which also illustrate a key attribute of such OA due to the SOI effect – Figure 1(d) illustrates this rotation for two eigenmodes, and it is evident that $Z_{beat}$ varies strongly with topological charge, since SOI is known to be strongly dependent on topological charge (varies as $L^3$ in the ring-core fibers used in our experiments, see Supplement 1). In addition, waveguide dispersion makes $Z_{beat}$ a strong function of wavelength too. Thus, a spatially complex but deterministic polarization rotation effect typically dependent primarily on the material (dispersive) properties of chiral media [16], is now dependent on light's topological charge itself [Fig. 1(f)]. Now, considering the Raman interaction, we can quantify the spatial frequency ($k = 2\pi/Z_{beat}$) difference of the linear polarization rotation between a pump and Stokes photon for arbitrary mode combinations as:

$$\Delta k = k_p - k_s = 2\pi \left[\frac{\Delta n_{eff}^{(p)}(\lambda_p)}{\lambda_p} - \frac{\Delta n_{eff}^{(s)}(\lambda_s)}{\lambda_s}\right] \qquad (3)$$

where, again, $p$ and $s$ represent pump and Stokes modes, respectively. When $\Delta k \neq 0$, which is the case for the majority of modes, Stokes light polarization walks off from the pump, see Fig. 1(d) – at every position $z$ past the input, the linear polarization orientation of pump (red) differs from that of Stokes light (green), leading to reduction in Raman scattering, since Raman gain maximizes for co-polarized light. In contrast, as Fig. 1(e) illustrates, polarization rotation rates, and hence $Z_{beat}$, are matched, and hence $\Delta k =$ 0, over narrow Stokes spectral ranges in the special case when Stokes light in the mode order $L_s = L_p - 1$ [denoted by the horizontal dashed lines of Fig. 1(f)]. In such a case, while Raman scattering is expected to occur over the entire spectral region depicted by the grey band [Fig. 1(f)], polarization evolutions are matched only at a specific wavelength within the grey Raman band, restricting Raman scattering to this region, where the $\Delta k = 0$ condition is satisfied. Note that this is reminiscent of phase matching conditions for parametric nonlinear (as opposed to phonon-assisted Raman) processes, which maximizes only over the (relatively narrow) spectral range in which phase matching or momentum conservation is achieved, even if the underlying nonlinear susceptibility is broadband.

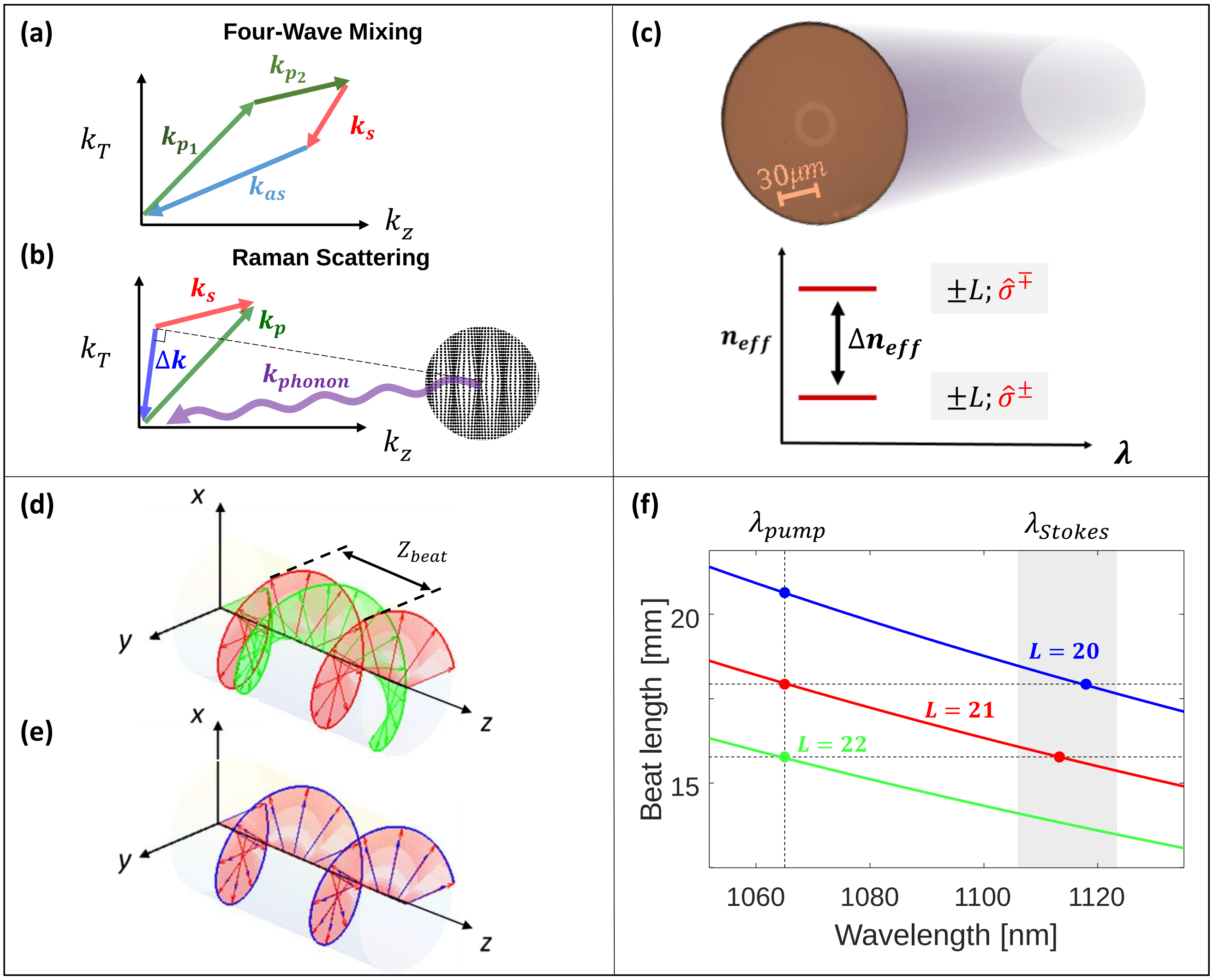

**Fig. 1. Phase-matched Stokes Raman scattering principle.** (**a**) Parametric nonlinear process such as four-wave-mixing requires momentum conservation of all interacting photons for maximum gain, arrows indicate wavevectors, and the subscript $p_1$, $p_2$, $s$ and $as$ represent two pumps, Stokes and anti-Stokes photons. (**b**) Stokes Raman scattering process is a phase-insensitive process as the optical phonons compensate for any phase mismatch between the pump and Stokes photons. (**c**) Facet image of the ring-core fiber used in the experiment, schematic plot of effective index splitting ($\Delta n_{eff}$) between $\hat{\sigma}^+$ and $\hat{\sigma}^-$ of same $L$ due to spin-orbit interaction. (**d, e**) Schematic plots of polarization evolution for optical activity modes of different topological charges at pump and Stokes wavelengths. Red: Pump light in $L_p = 21$, green: Stokes light in $L_s = 22$, blue: Stokes light in $L_s = 20$. (**f**) Beat length as a function of wavelength for different modes.

## Results:

We probe this effect with a 650-ps, 1.7-kW-peak-power ($P_{peak}$) pump source comprising a Nd:YAG laser and a fiber amplifier. This pump is converted to the desired OAM mode of a 9.8-m-long ring-core fiber [Fig. 1(c)] with a spatial light modulator (SLM), and another SLM is used to modally sort the output spectrum by the topological charge (see Supplement 1). We sweep $L_p$ from 7 to 24 in both $\hat{\sigma}^-$ and $\hat{x}$ (OA) polarizations whose resultant measured mode intensity and phase profiles at the fiber output are shown in Fig. 2(b) (the details of the fiber and mode properties are discussed in Supplement 1). Similar ring shapes and sizes for all the modes result in similar nonlinear overlap integrals $\eta$ [see Eq. (2)] for all different combinations of $L_p$'s and $L_s$'s – which is what conventionlly controls the Raman scattering strength. To account for the slight differences, and hence to avoid any variations in Raman scattering strengths due to differing pump-Stokes overlaps, we adjust $P_{peak}$ slightly for each $L_p$ such that $(P_{peak} \cdot \eta)$ remains constant throughout the experiments. For pumps in a pure circular polarization $\hat{\sigma}^-$, we observe that the Stokes light is also in $\hat{\sigma}^-$, and almost identical, conventionally co-polarized Raman spectra are obtained for all pump and Stokes mode combinations [17]. Given their similarity, for illustrative clarity, we only plot one of them ($L_p = 7, \hat{\sigma}^-; L_s = 8, \hat{\sigma}^-$) as the red curves in Fig. 2(a). But when the pump is linearly polarized along $\hat{x}$, despite the fact that the spatial overlap, $\eta$ still remains similar across all mode combinations, the non-zero $\Delta k$ terms for Stokes and pump combinations satisfying $(L_s \neq L_p - 1)$ result in complete polarization walk-off between the pump and Stokes light [shown schematically in Fig. 1(d)], leading to significant Raman gain reduction. Again, since all $(L_s \neq L_p - 1)$ yield similar spectra, Fig. 2(a) plots the spectra obtained for only one such mode combination – $L_p = 7, \hat{x}$ and $L_s = 8$ [green curves in Fig. 2(a)]. Note that while their spectral shapes are similar to the conventional case where the pump is in $\hat{\sigma}^{\pm}$ polarization, power in the Raman band is dramatically suppressed by ~20 dB for the *same* pump powers and mode sizes (intensity overlaps) [18]. Finally, the blue curves in Fig. 2(a) are Raman spectra for the Stokes mode in $L_s = L_p - 1$. Due to the perfect polarization overlap at every position along the fiber [Fig. 1(e)], a strong Raman gain peak with strength similar to that of co-polarized Raman scattering (red curves) is obtained at the beat length matching wavelength $\lambda_b$ where $\Delta k = 0$. Furthermore, the Raman spectrum is re-shaped and narrowed (yellow shaded region under blue curve) since $\Delta k$ increases as Stokes wavelength deviates from $\lambda_b$. Additional sharp peaks evident on every spectral plot arise from the multitude of parametric FWM interactions that are also possible in these fibers [19] – as described in Supplement 1, these neither influence, nor change the conclusions evident from the Raman spectra. The modulated Raman peak wavelength $\lambda_b$ shifts as pump mode changes since the $\Delta k = 0$ condition is strongly dependent on the spectral dispersion of the modes, as well as, crucially, the topological charge that the pump mode carries. Hence, Fig. 2(a) illustrates the ability to tailor both the strength and wavelength of Raman emission by controlling the angular momentum content of light.

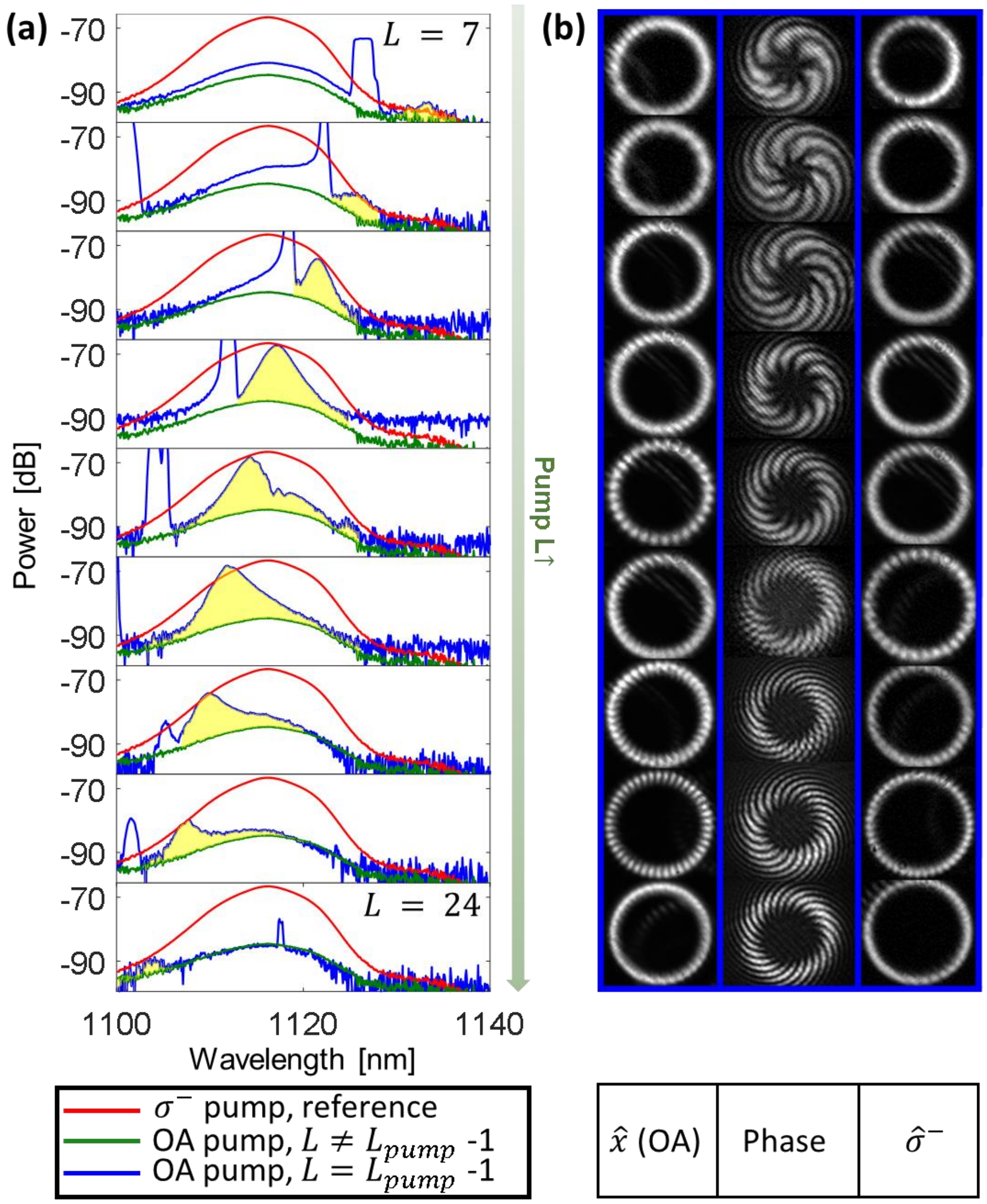

**Fig. 2. Raman spectra and pump mode images.** (**a**) Raman spectra plots for pump with different topological charges, arranged, top to bottom as $L_p = 7, 8, 9, 11, 14, 17, 19, 21, 24$; Raman spectra shown as red curves represent {Stokes in $L_s = 8$; Pump in $L_p = 7, \hat{\sigma}^-$} combinations; green curves represent {$L_s = 8$; $L_p = 7, \hat{x}$} combinations; and blue curves denote {$L_s = L_p - 1$; $L_p, \hat{x}$} combinations. The yellow shaded area represents contributions from the $\Delta k = 0$ phase matching condition. (**b**) Left and right columns: Intensity distribution of pump modes in $\hat{x}$ and $\hat{\sigma}^-$ polarizations. Middle: Phase pattern acquire by interference with an expanded Gaussian beam, the number of parastiches in the spiral patterns indicating $L_p$, the order of OAM of the pump beam.

Figure 3(a) shows the integrated Raman power for the three functionally distinct mode combinations described in Fig. 2. The red curve represents the co-polarized Raman scattering process where both pump and Stokes light are in the $\hat{\sigma}^-$ polarization, with each data point and error bar denoting the average and deviation of Raman strength in all Stokes modes for a given $L_p$. Despite the fact that the error bar for the $L_p = 11$ case is large, due mainly to superimposed spectra from parasitic FWM (see Supplement 1), the flat trend with respect to $L_p$ confirms conventional wisdom, that Raman strength does not depend on the phases of the participating waves (note that topological charges essentially manifest in the spatial phase of a mode). In contrast, when the pump light is an optically active state arising from SOI interactions, all but one of the Stokes modes experience phase mismatches due to non-zero $\Delta k$, and Raman power is suppressed by ~ 20 dB, as illustrated with the green curve of Fig. 3(a) (again, large measurement errors for $L_p = 9, 12$ arise from parasitic FWM, but do not, otherwise, influence the observed trend). For the Stokes mode corresponding to $L_s = L_p - 1$, we observe a systematic tunability of Raman strength over 15 dB since the Raman spectra are now tailored by the $\Delta k = 0$ condition [blue curve in Fig. 3(a)]. The spectral position of peak Raman gain now depends not only on the conventional Raman scattering strength of the material, but also its relation to the beat-length-matched wavelength $\lambda_b$. Figure 3(b) shows this new degree of freedom – the ability to tune the Raman gain peak wavelength $\lambda_b$, simply by tuning the topological charge of the pump mode, by over 30 nm (~ 8 THz), which is greater than half of the material's (Silica's) conventional Raman Stokes shift (~13 THz).

Further affirmation of this phase-matched behavior is evident from recording the bandwidth of the modified Raman spectra [yellow shaded regions of Fig. 2(a)] as a function of fiber length. The spectra of self-phase matched, conventional Raman scattering should display no inherent dependence on interaction length, at least in the spontaneous scattering regime. In contrast, processes that have strict phase matching requirements, such as second harmonic generation, have a response that is proportional to both phase mismatch $\Delta k$ and propagation distance $z$:

$$I \propto sinc^2\left(\frac{\Delta k \cdot z}{2}\right) \quad (4)$$

Since $\Delta k$ maps to frequency, the bandwidth of a parametric process is inversely proportional to $z$. Figure 3(c) plots the simulated $\Delta k \cdot z/2$ values and Fig. 3(d) shows the corresponding experimentally recorded Raman spectra of $L_p = 15$; $L_S = 14$, for two different fiber lengths – 9.8m and 20m – shown in blue and red curves, respectively. The green band in the simulations indicate the spectral limits of $\Delta k \cdot z/2$ at which the nonlinear response is halved [per Eq. (4)]. Remarkably, not only do these intersection points correspond well with the measured Raman spectra for the two lengths, the measured bandwidth ratio (1.42 nm/2.64 nm) also closely corresponds to the fiber length ratio (9.8 m/20 m). This length dependence of an otherwise self-momentum conserved scattering process illustrates the centrality of the role of phase matching arising from SOI interactions.

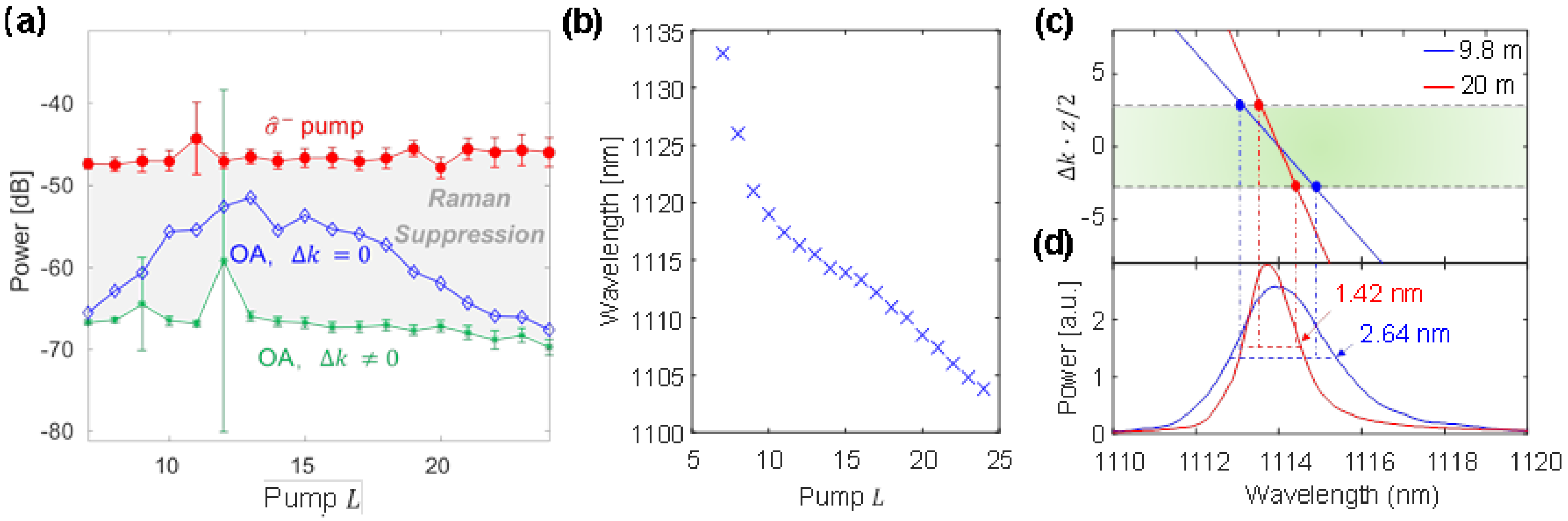

**Fig. 3.** (**a**) Integrated Raman power within 20 nm wavelength range for different pump-Stokes mode combinations, red: Raman power for {Stokes in $L_s, \hat{\sigma}^-$; Pump in $L_p, \hat{\sigma}^-$}; green: Raman power for $\{L_s \neq L_p - 1; L_p, \hat{x}\}$ combination, i.e. $\Delta k \neq 0$; blue: Raman power for $\{L_s = L_p - 1; L_p, \hat{x}\}$ combination, i.e. $\Delta k = 0$. Relatively larger error bars for some datapoints in red and green curves represent cases where a coincident, parasitic FWM peak existed, corrupting measurement of Raman power alone. (**b**) Peak wavelengths shift for Stokes mode in $L_s = L_p - 1$ as a function of pump mode $\{L_p, \hat{x}\}$. (**c**) Simulated plot of phase mismatch multiplied by fiber length as a function of wavelength for $\{L_s = 14; L_p = 15, \hat{x}\}$ mode combination. (**d**) Experimentally measured Stokes Raman spectral shape for $\{L_s = 14; L_p = 15, \hat{x}\}$ mode combination in both 9.8 and 20-m-long ring-core fibers.

## Discussion:

The ability to modulate Raman scattering strength by ~20 dB and to spectrally tune it by over ½ the Raman gain bandwidth by controlling light's topological charge yields a design degree of freedom to enhance, spectrally shape, or avoid Raman scattering, as applications as diverse as spectroscopy, lasers or quantum sources may demand. The fact that the phase of light plays a role opens this important nonlinear process to control via dispersion engineering – an established design tool for fibers and waveguides. Finally, since the underlying effect arises from SOI of topologically complex light, systems other than waveguides, such as tightly focused light in bulk media, or OAM beams in chiral media, which would experience SOI or optical activity, may yield similarly anomalous Raman scattering dependencies, enabling control of Raman scattering in media other than waveguides too.

**<u>Supplementary Materials:</u>** Systematic control of Raman scattering with topologically induced chirality of light

## Materials and methods

### Detailed experiment setup

Figure S1 illustrates the experimental setup: We use a homemade laser source that comprises a Nd:YAG seed laser and 1.5-m-long Yb-doped fiber amplifier that produces 0.65 ns, up to 120 kW peak power, pulses at the repetition rate of 19.5 kHz. The attenuator consists of a half wave plate and polarization beam splitter, which serves the purpose of controlling power and ensuring output light in $\hat{x}$ polarization, which is the preferred polarization for the spatial light modulator (SLM). The input Gaussian beam is then converted into the desired OAM state with topological charge $L$ using a SLM encoded with spiral phase, and the spin angular momentum (SAM) of light is controlled with a quarter wave plate (QWP). In our experiment, we specifically compare $\hat{\sigma}^-$ and $\hat{x}$-polarized OAM states. The free space OAM beam is then coupled into the specially designed ring-core fiber that supports stable propagation of multiple OAM beams, a detailed analysis of the fiber and modes properties is presented in §1. At the output, the polarization control element varies for the two different pump polarization cases: (1) for the experiments in which the pump is in the $\hat{x}$ polarization, it only comprises a linear polarizer (LP) whose axis is sequentially set to horizontal and vertical positions for interrogating the output nonlinear spectra in $\hat{x}$ and $\hat{y}$ polarizations, respectively; (2) for the cases where the pump is in the $\hat{\sigma}^-$ polarization, we add a QWP, and the combination of the QWP and LP acts as a circular polarizer, which allows us to measure light in $\hat{\sigma}^\pm$ polarizations separately. We use a second SLM and a single mode pick up fiber (part number SM980, called just SMF, henceforth) to modally sort the spectrum by its topological charge. The basic principle of the mode sorting function can be understood as follows: the OAM beam from the output of the fiber, with topological charge $L_1$, is incident on the mode sorter SLM which has another spiral phase pattern of OAM value $L_2$. The resultant beam, upon reflection from this SLM, has an OAM value of $\Delta L = L_1 - L_2$. For all input beams resulting in non-zero $\Delta L$, light remains in a donut-shaped OAM mode in free space (see inset of Fig. S1) and cannot be coupled into the single mode fiber. In contrast, for $L_1 = L_2$, the OAM gets cancelled and the output is converted to a Bessel beam with *zero* OAM in free space, which can now couple into the SMF. Hence, the mode sorter SLM and SMF combination can selectively measure spectra only in $L_2$. Furthermore, by sweeping $L_2$ and obtaining the spectra from each orthogonal polarization, we are able to measure the total output spectra in each individual mode.

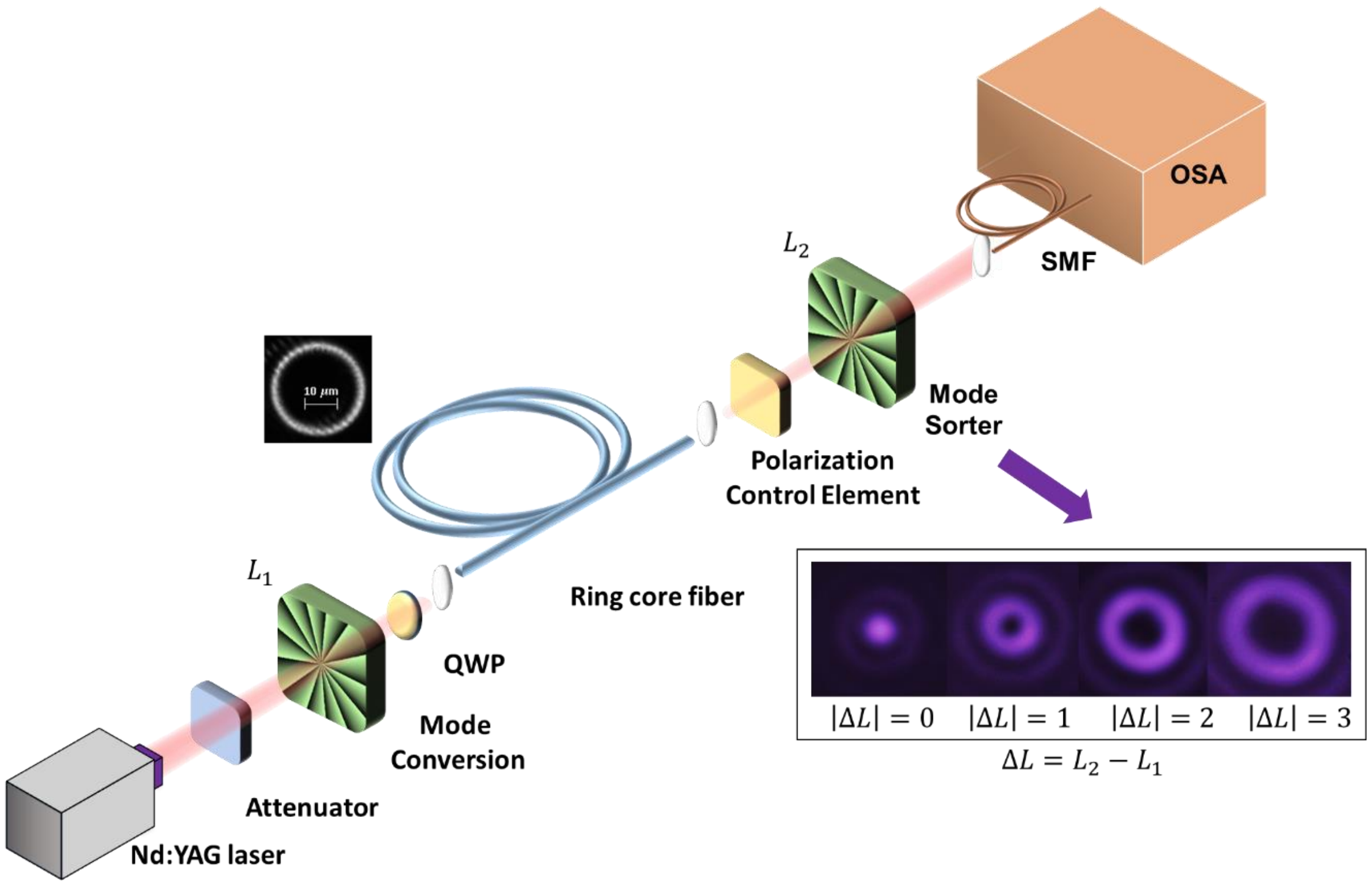

**Fig. S1.** Detailed experiment setup, the attenuator is comprised of a half wave plate and polarization beam splitter, mode conversion and mode sorter are two identical SLMs with different spiral phase patterns, the polarization control element varies from a single LP and a combination of QWP and LP depending on the pump polarizations, details discussed in §1. The inset picture shows the free-space reconverted beam shape for different OAM values of two SLMs. SMF: single mode fiber, OSA: optical spectrum analyzer.

## Mode sorter setup characterization

Since the extinction ratio of the mode sorter is finite, the spectral contribution of one mode may appear in other modes as well, which is especially problematic for weeding out contributions from parasitic FWM processes, which are typically intense. Figure. S2A shows the mode-sorted nonlinear spectra for the participating modes when pumping in an $L = 23$, OA state. The sharp peak showing up at 1119.52 nm is expected to be due to FWM, and should reside in the $L = 21$ mode according to simulated phase matching conditions (discussed in more detail in §2). The experiment confirms this prediction – see yellow curve in Fig. S2A. However, due to the limited extinction ratio of the mode sorter, we also get the projection of this spectrum in the other modes, necessitating the need for calibration of our mode-sorter, based on the a priori knowledge that the FWM spectrum should have resided only in a specific mode of given $L$. We set up the relationship between the actual mode power $P$ and measured power $P'$ at the output of SMF through the efficiency matrix in Eq. S1, where each element $\eta_{m,n}$ in the matrix stands for the coupling efficiency for input light in $L_m$ when SLM spiral phase pattern is set to be $L_n$,

$$\begin{bmatrix} P'i \\ P'_{i+1} \\ \vdots \\ P'_j \end{bmatrix} = \begin{bmatrix} \eta_{i,i} & \eta_{i+1,i} & \cdots & \eta_{j,i} \\ \eta_{i,i+1} & \eta_{i+1,i+1} & \cdots & \eta_{j,i+1} \\ \vdots & \vdots & \ddots & \vdots \\ \eta_{i,j} & \eta_{i+1,j} & \cdots & \eta_{j,j} \end{bmatrix} \cdot \begin{bmatrix} Pi \\ P_{i+1} \\ \vdots \\ P_j \end{bmatrix} \tag{S1}$$

To characterize all the elements in the efficiency matrix, a pure OAM state input is required. Nonlinearly generated OAM modes due to FWM are excellent candidates in this regard since phase matching and OAM conservation conditions ensure single mode emission at the phase-matched wavelength [19]. Each column of the matrix is acquired by measuring coupling efficiency for input light, expected to be purely in $L_m$, and sweeping SLM phase pattern $L_n$ from $L_m - 4$ to $L_m + 4$. Probing spectra in modes $|L_m - L_n| > 4$ is not necessary since the coupling efficiency is negligibly small ($< -40$ dB). By sweeping $L_m$ and repeating the measurements, we can construct the whole efficiency matrix. We perform an independent calibration matrix measurement by sweeping $L_m$ from 5 to 28, since the power in neighboring modes is very sensitive to change of $L_m$ for extinction ratios greater than -15 dB. We take the average for all the extinction ratio measurements and create an average efficiency matrix (Fig. S2C).

Hence, we conclude that the extinction ratio is -17 dB for $\Delta L = \pm 1$, -24 dB for $\Delta L = \pm 2$, -30 dB for $\Delta L = \pm 3$. We then apply the efficiency matrix to the measurement by Eq. S1 to further improve the measurement accuracy, as is shown in Fig. S2B. As is evident, peaks at 1192.52 nm in the unexpected modes are substantially suppressed compared to those in Fig. S2A, allowing for precise measurements of the modal content of output spectra.

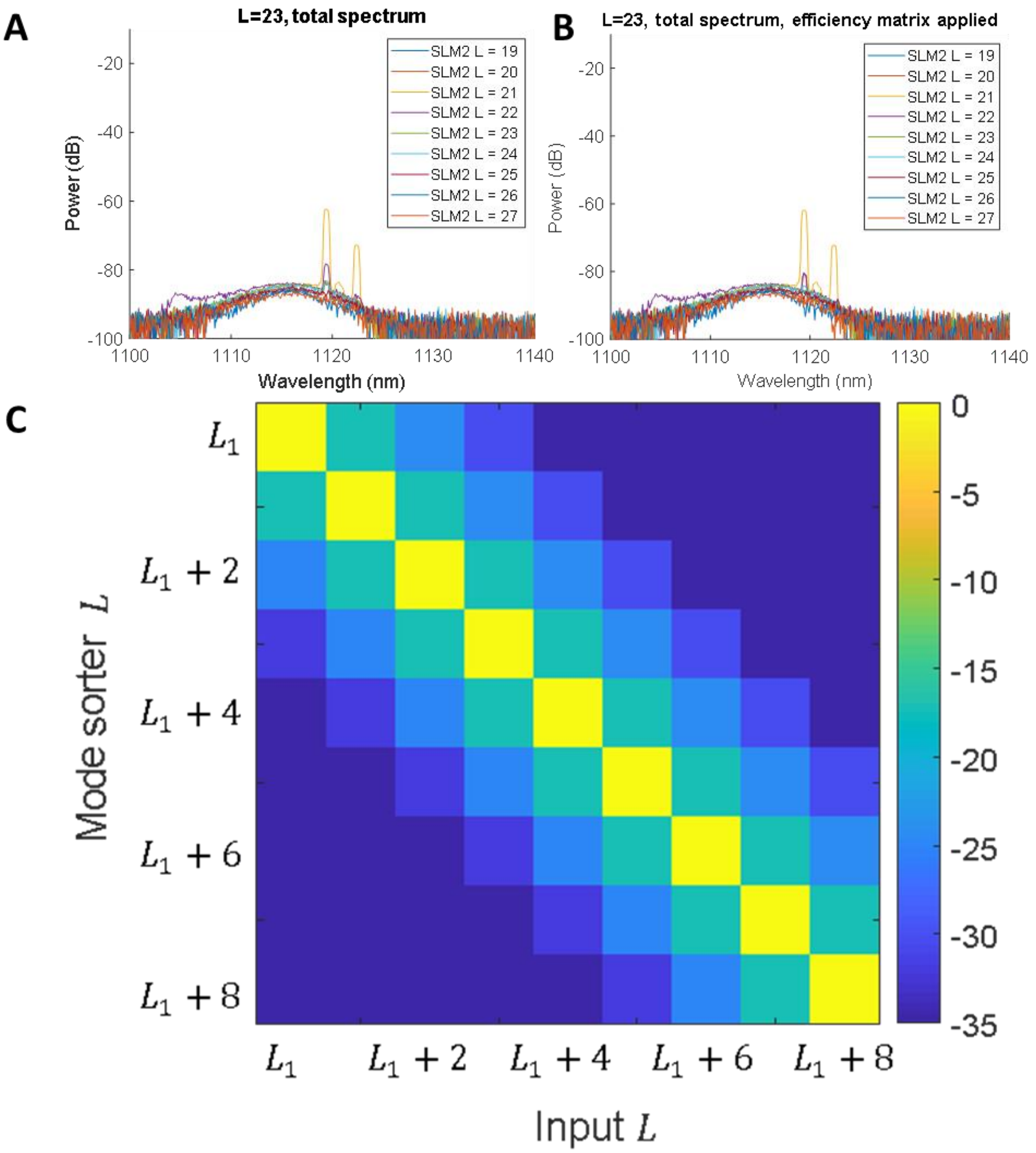

**Fig. S2. (A)** Mode sorted spectrum for pump in $L = 23$, OA state without applying the efficiency matrix, sharp peaks at ~ 1120 nm correspond to FWM Stokes peak in $L = 22$. **(B)** Mode sorted spectrum for pump in $L = 23$, OA state after applying the efficiency matrix per Eq. S1. **(C)** Calibrated efficiency matrix, each column corresponds to the coupling efficiency when sweeping the mode sorter OAM value, the noise floor of the system is set to be -35 dB, which is restricted by the detector noise floor.

## Supplementary Text

### 1) Fiber design

#### 1a) Fiber profile and effective index of guided modes

The refractive index profile of the ring-core fiber used in our experiments is measured with a commercial interferometer-based fiber index profiler (Interfiber Analysis: IFA-100), and the refractive index $\Delta n(r)$ relative to Silica index is shown in Fig. S3A. The OAM modes are guided in the ring structure region whose inner diameter and thickness are 21 $\mu\mathrm{m}$ and 5 $\mu\mathrm{m}$, respectively. The large refractive index step of 0.038 evident in Fig. S3A enables a large mode volume as well as significant separation of the effective index $n_{eff}$ (related to the propagation constant $\beta$ of a mode by $n_{eff} = \beta\lambda/2\pi$), among modes. This separation $\Delta n_{eff}$, especially between modes of same $|L|$ but different combinations of OAM and SAM signs (called the spin-orbit aligned (SOa) modes when $L$ and $\hat{\sigma}$ have the same sign and spin-orbit anti-aligned (SOaa) modes when they have opposite signs) is a key feature ensuring stability of not only each individual OAM mode[16]. The mode simulation is performed by solving the full-vectorial wave equation with a finite-difference scheme [21] and the wavelength dependent effective indices are presented in Fig. 3B. Each color represents modes with different topological charges and is comprised of two curves for SOa-SOaa pairs of same $L$. As is shown in the inset figure in Fig. 3B, the solid purple line corresponds to $L = 19, SOaa$ state and the dashed purple line corresponds to $L = 19, SOa$ state, the separation between these two modes is usually on the order of $10^{-5}$, which is orders of magnitude smaller than $\Delta n_{eff}$ between different $L$'s (~$10^{-3}$ as indicated by the blue arrows), hence it looks like they overlap with each other in Fig. 3B. $\Delta n_{eff}$ between these pairs of non-degenerate modes arises from spin-orbit interactions and usually increases monotonically with $L$ (discussed in more detail in §1b). From our experience, stable guidance of each mode from these pairs is ensured for $\Delta n_{eff}$ ~ $5 \times 10^{-5}$, especially for the meter-long fibers used in our experiments. This would suggest that the lowest mode order stably guided in this fiber is $L = 21$. However, experimentally, we found that even lower order modes are also guided stably – an anomalous result which we discuss in greater details in §1b. The highest (cutoff) mode is determined by the relative position of the $n_{eff}$ curve to the refractive index of Silica at that wavelength (calculated from Silica's Sellmeier equation and plotted as the black dashed curve in Fig, 3B). Conventionally, any $n_{eff}$ curve that is below the Silica index curve is considered to be beyond cutoff, hence in this case, $L = 25$ is expected to be the highest mode order supported by the fiber at the pump wavelength at 1064 nm.

## 1b) Spin-orbit interaction (SOI)

A spin-orbit interaction in fibers arises from the anisotropy of the fiber's refractive index distribution. It serves to split the $n_{eff}$ for two circularly polarized modes of same *L* by [22]:

$$\Delta n_{eff} = \frac{\lambda^2 L}{4\pi^2 a^2 n_{co}^2} \int_0^\infty |E(r)|^2 \frac{\partial \Delta n(r)}{\partial r} dr \tag{S2}$$

where $r$ is the radial coordinate, $E(r)$ is the normalized electric field for the unperturbed mode, $a$ is the size of the fiber core, $\lambda$ is the free-space wavelength, and $n_{co}$ is the maximum refractive index. Note the similarity of this equation with SOI in atomic systems, where eigenstate splitting increases with *L.* The ring-core design of Fig. S3A is among a class of fiber designs that exacerbates this splitting, which, in turn, ensures that each of the modes propagates in these fibers stably[16].

The exact dependence of $\Delta n_{eff}$ on *L* is influenced by how the field $E(r)$ varies with *L* [23]. For the fiber shown in Fig. S3A, $\Delta n_{eff}$ scales as $L^3$, as evident from Fig. S3C, which plots the simulated $\Delta n_{eff}$ versus $L^3$ in blue as well as its excellent match with a fit of a function $y = constant * L^3$ (shown in green, linear regression $R^2$ coefficient 0.999). To experimentally verify this relationship, we measure $\Delta n_{eff}$ values for each $L$ based on the linear polarization rotation angle $\theta_{OA}$ of their corresponding OA state, which is defined as[6]:

$$\theta_{OA} = \frac{\pi \Delta n_{eff}}{\lambda} d \quad \text{(S3)}$$

where $d$ represents the propagation distance and can be accurately measured via cutback. Hence, we can calculate $\Delta n_{eff}$ by launching an $\hat{x}$-polarized (OA) light into the fiber and measuring $\theta_{OA}$ at the output with an LP. To avoid the issue of uncertainties arising from unwrapping the measured phase angles, our cutback length *d*, were set to be less than 1 cm, hence ensuring $\theta_{OA} \epsilon [0, \pi]$ for all the $L$'s. The measured $\Delta n_{eff}$ vs. $L^3$ is plotted as the red circles in Fig. 3C. While the measurements agree well the simulation results for modes of higher *L,* the mismatch gets larger as $L$ decreases. The origin of this discrepancy is likely from an incorrect measurement of $\Delta n(r)$ on two accounts. Firstly, no refractive index profiler would have arbitrary spatial resolution, and as Eq. S2 shows, $\Delta n_{eff}$ is sensitive to the *gradient* of this measurement. Secondly, and perhaps more importantly, we believe that the anomalously large (but opposite in sign) $\Delta n_{eff}$ one experimentally observes for lower *L* modes arises from the fact that low *L* modes have larger field amplitudes closer to the inner ring-core boundary, and large stresses and strains due to the large $\Delta n(r)$ at this boundary result in inherent radial or azimuthal birefringence. That is, $\Delta n(r)$ is now a vector quantity and the spin-orbit splitting cannot be simply represented by Eq. S2. This is speculation (by elimination of other possible causes), of course, and confirming the origins of the large $\Delta n_{eff}$ for low *L* modes will require more in-depth measurement of the permittivity tensor [and not just the scalar $\Delta n(r)$] of the fiber profile.

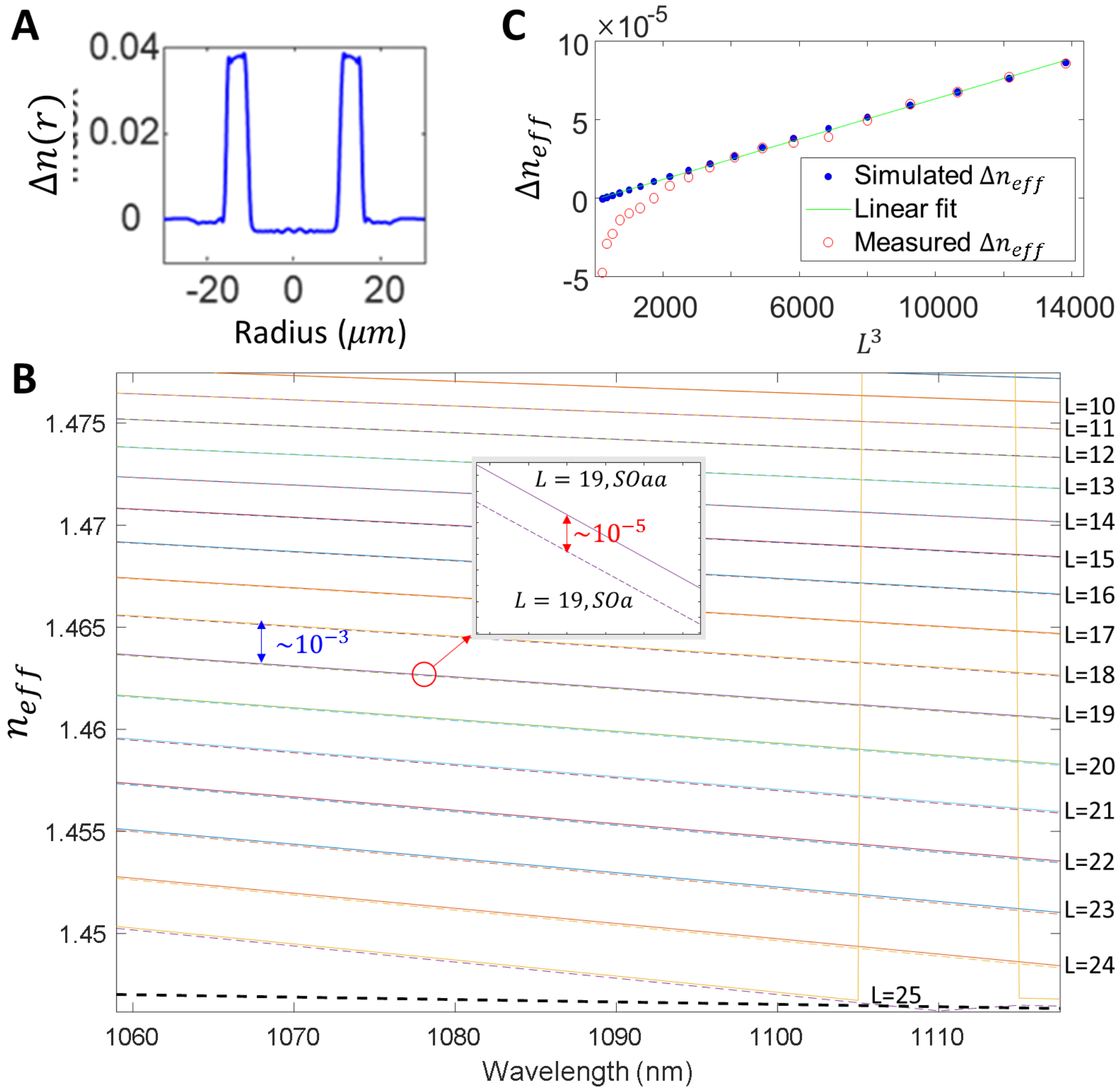

**Fig. S3. (A)** Refractive index $\Delta n(r)$ of the ring-core fiber used in the experiment, the index profile is relative to the index of Silica, which is the material that forms the cladding of the fiber. **(B)** Effective index ($n_{eff}$) for all the OAM modes as a function of wavelength, the black dashed line corresponds to the refractive index of Silica calculated by the Sellmier equation. The inset figure shows the zoomed-in $n_{eff}$ in the red circle region. $n_{eff}$ curves for each $L$ consist of a solid and dashed line, representing, respectively, the SOaa and SOa modes. **(C)** Spin-orbit interaction (SOI) induced splitting $\Delta n_{eff}$ of SOa-SOaa mode-pairs for the variety of OAM ($|L|$) orders probed in our experiments. Blue: Simulated $\Delta n_{eff}$ as a function of $L^3$, green: linear fit of $\Delta n_{eff}$ vs. $L^3$, red: measured $\Delta n_{eff}$ as a function of $L^3$.

2) Parasitic Four Wave Mixing process

As mentioned in the context of Fig. 2A, the additional sharp spectral features shown on top of the blue curves result from FWM among different OAM modes. They are unavoidably present, given the diversity of phase matching conditions arising from the plethora of available modes. Their emission wavelengths are determined by the conventional phase matching condition [20]:

$$\Delta\beta = \frac{n_{eff}^{p}(\lambda_p)}{\lambda_p} + \frac{n_{eff}^{q}(\lambda_q)}{\lambda_q} - \frac{n_{eff}^{s}(\lambda_s)}{\lambda_s} - \frac{n_{eff}^{as}(\lambda_{as})}{\lambda_{as}} = 0 \quad \text{(S4)}$$

where $p, q, s$ and $as$ represent the first and second pumps, Stokes and anti-Stokes beams, respectively. This nonlinear process must also conserve angular momentum, and so:

$$-L_p - L_q + L_s + L_{as} = 0 \quad \text{(S5)}$$

Hence strong emissions happen only when both the residual topological charge is equal to 0 and the phase matching condition is fulfilled. In our experiment, we usually observe two FWM processes:

$$L_s = L_p + 1 \leftarrow L_p + L_p \rightarrow L_{as} = L_p - 1 \quad \text{(S6)}$$

$$L_s = L_p + 2 \leftarrow L_p + L_p \rightarrow L_{as} = L_p - 2 \quad \text{(S7)}$$

here we refer to Eq. S6 and S7 as $\Delta L = 1$ and $\Delta L = 2$ FWM processes, where the pumps are degenerate in topological charge $L_p = L_q$, anti-Stokes light is one or two mode orders lower than that of the pump, and hence according to Eq. S4, Stokes light is in one or two mode orders higher. Combining Eqs. S4-7, we can simulate the Stokes light wavelengths and compare with experiment – as is shown in Fig. S4A, we find excellent agreement, with wavelength deviations to be less than 3 nm for any of the FWM processes. When the FWM Stokes wavelength falls into the grey band (see Fig. 2A) where material Raman gain $g_R(\Omega)$ is substantial, the spectra due to Stokes Raman scattering and FWM become hard to distinguish, with the FWM peaks showing up as sharp features on top of the broad Raman peak. These extraneous peaks can be easily excluded when calculating integrated Raman power in most cases, since they are relatively narrowband compared to the (Raman) spectra of interest to us. However, in two special cases, specifically, when the pump is in the $L = 9$; OA or $L = 12$; OA modes, we also observed broadband FWM emissions on top of Raman (see Fig. S4B and S4C), which correspond to degenerate $L$ FWM process where all the interacting photons are in the same $L$. Since the bandwidth is almost similar to the Raman gain peak, large error bars result in integrated Raman power at those data points in Fig. 3A. We had argued that the large error bars may be ignored because the rest of the data points do indeed follow the trend we observe – below, in §3b, we offer additional assurance for the validity of this assumption.

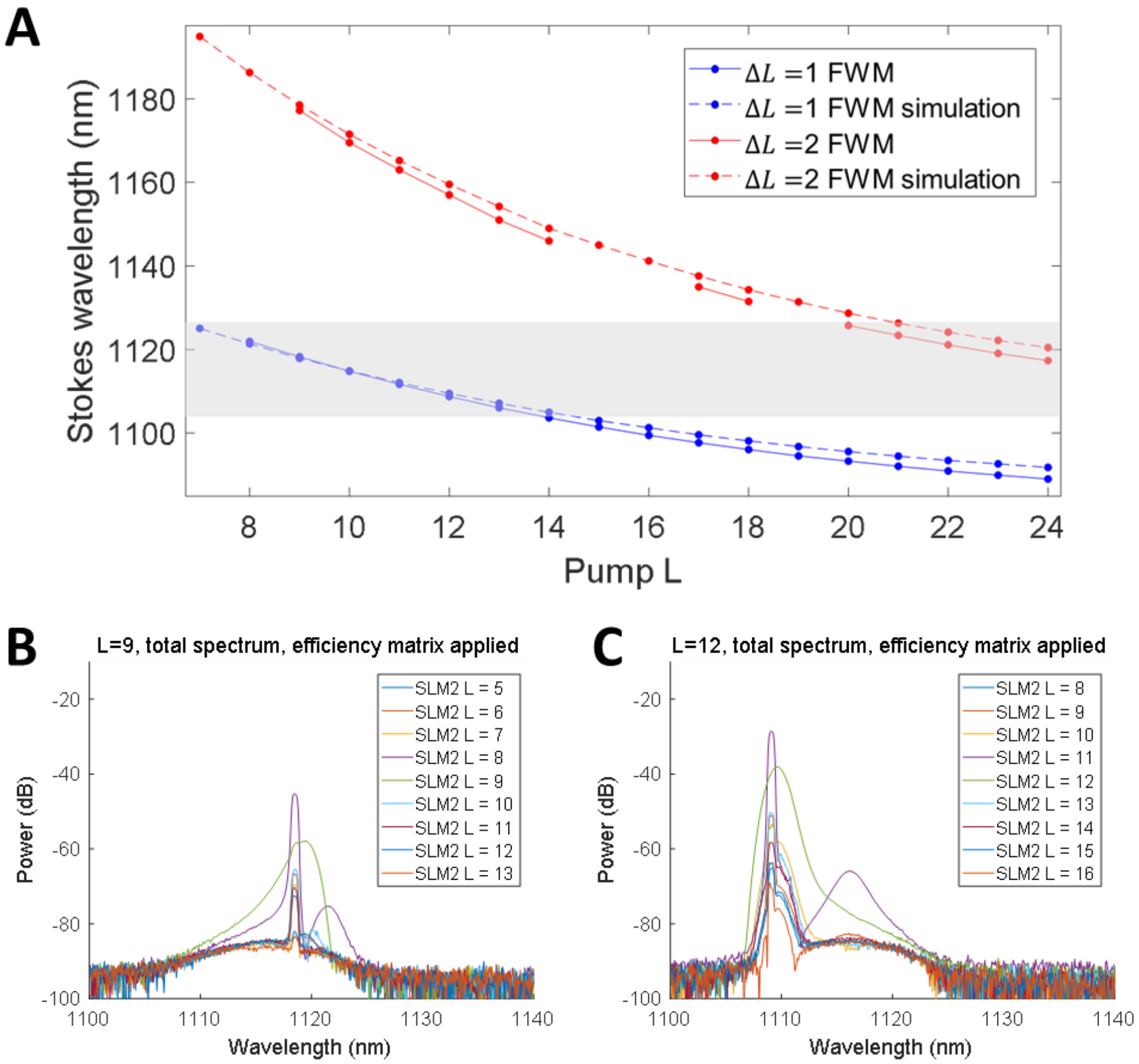

**Fig. S4. (A)** Stokes wavelength for two different types of four wave mixing process, red curves correspond to $\Delta L = 1$ FWM process where the Stokes and anti-Stokes mode are 1 mode order away from the pump, and blue curves correspond to $\Delta L = 2$ FWM process. Dashed lines are simulation results and solid lines are experimentally measured. **(B)** Mode sorted spectra when pumping in $L = 9$, OA state, a broad band FWM peak shows up at ~1120 nm in the same mode as the pump $L = 9$. **(C)** Mode sorted spectra when pumping in $L = 12$, OA state, similarly, a broad band FWM peak shows up at ~1110 nm in the same mode as the pump $L = 12$.

## 3) Multimode generalized nonlinear Schrödinger equation simulation

### 3a) Phase matching behavior of topological charge mediated Raman scattering process

In order to verify the phase dependence of the optical activity mediated Raman process, we perform simulations by numerically solving the multimode generalized nonlinear Schrödinger equation[24], which can be written in the following form: (Shock term ignored for quasi-CW pulses)

$$\frac{\partial A_p(z,t)}{\partial z} = i\left(\beta_0^{(p)} - \beta_0\right) A_p(z,t) - \left(\beta_1^{(p)} - \beta_1\right)\frac{\partial A_p(z,t)}{\partial t} + i\sum_{n\geq 2}\frac{\beta_n^{(p)}}{n!}\left(i\frac{\partial}{\partial t}\right)^n A_p(z,t)$$

$$+\frac{in_2\omega_0}{c}\sum_{l,m,n}\left\{Q_{plmn}^{(1)}(\omega_0)2A_l(z,t)\int d\tau R(\tau)A_m(z,t-\tau)A_n^*(z,t-\tau)\right.$$

$$\left.+\,Q_{plmn}^{(2)}(\omega_0)A_l^*(z,t)\int d\tau R(\tau)A_m(z,t-\tau)A_n(z,t-\tau)e^{2i\omega_0\tau}\right\} \qquad \text{(S8)}$$

The overlap integrals can be written in the following form:

$$Q_{plmn}^{(1)} = \frac{\varepsilon_0^2 n_0^2 c^2}{12}\int dxdy\left[F_p^*(\omega)\cdot F_l(\omega)\right]\left[F_m(\omega)\cdot F_n^*(\omega)\right] \qquad \text{(S9)}$$

$$Q_{plmn}^{(2)} = \frac{\varepsilon_0^2 n_0^2 c^2}{12}\int dxdy\left[F_p^*(\omega)\cdot F_l^*(\omega)\right]\left[F_m(\omega)\cdot F_n(\omega)\right] \qquad \text{(S10)}$$

where $F$ represents the normalized radial distribution of the mode electric field, $R(\tau)$ stands for the nonlinear response function, and can be written as:

$$R(\tau) = (1-f_R)\delta(t) + \frac{3}{2}f_R h(t) \qquad \text{(S11)}$$

where $f_R = 0.18$ is the fractional contribution of the Raman response to the total nonlinearity and $h(t)$ is the delayed Raman response function, modeled after the experimental Raman response function of silica [24], and the delta function corresponds to Kerr effect. We first involve only 6 OAM modes with 3 different topological charge values 14, 15, and 16 of both circular polarizations $\hat{\sigma}^{\pm}$. The input power and fiber length are the same as in the experiments, i.e., 1.7 kW and 9.8 m. We assume a quasi-CW pulse, and quantum shot noise is included as one photon with random phase per longitudinal mode (frequency bin) per OAM mode. Each spectrum shown is an average of 50 simulations with different random number generator seeds.

We start our simulation by pumping in only the $L = 16, \hat{\sigma}^+$ mode. Note that for this specific mode combination, FWM among modes with different topological charges is prohibited because the pump mode has the highest $L$, and so scattering into $L = 14\ or\ 15$ would require a higher-order mode to be present according to the OAM conservation rule (Eq. S5). These higher-order modes have intentionally been omitted from the simulations for the purpose of studying the Raman effect without the presence of FWM. The comparison where FWM is included is discussed in §3b. The resultant spectra look very similar for all the $L$'s in the same polarization as that of the pump ($\hat{\sigma}^+$), and we just select one mode (Stokes in $L = 15$) and show as the red curve in Fig. S5A, which represents the conventional co-polarized Raman gain spectrum. We then switch the pump to an OA state of the same $L = 16$ while holding total power constant. As is evident from the dark and light green curves in Fig. S5A, Stokes light in $L = 14$ and $L = 16$ are highly suppressed because they cannot fulfill the beat length matching condition across the Raman gain band, as illustrated in Fig. 1D. In contrast, for Stokes light in $L = 15$, we observe a narrow modulated Raman spectral shape that peaks at ~1110 nm, which is determined by the beat length matching condition, as is also shown experimentally in Fig. 2A. To further support the idea of controlling Raman spectral shape by dispersion engineering, we keep $P \cdot \eta$ constant and sweep pump $L$ to acquire the Raman spectra in the beat-length-matched mode, as shown in Fig. S5C. The corresponding beat length matching wavelength $\lambda_b$ is plotted in Fig. S5D. The good match between simulated $\lambda_b$ and the modulated Raman gain peak (indicated by the black dashed vertical lines) and analogous behavior shown in the experiments depicted in Fig. 2 supports our primary claim – that this topological effect is a manifestation of a heretofore unobserved phase matching behavior for Raman scattering.

3b) Four-wave-mixing effect on Raman scattering process

To probe whether the parasitic FWM peaks in our experiments affect, or otherwise modify, the phase-matched Raman scattering process, we simulate the behavior by intentionally turning on or off FWM. Above (§3a), we simulated the behavior by effectively turning off FWM. Now, we turn the effect on by replacing the $L = 14$ mode with the $L = 17$ mode in the ensemble of participating modes included in our simulation. The new choice of interacting modes are now $L = 15, 16 \text{ and } 17$. For pump in $L = 16$, $\Delta L = 1$ FWM (Eq. S6) processes are now allowed as both $L = 15$ and 17 are included in the simulation. As evident from Fig. S5B, when pumping in a single mode in $L = 16, \hat{\sigma}^{+}$, we obtain a similar conventional Raman scattering with FWM Stokes peak showing up at 1100 nm in $L = 15$ (red curve). Changing the pump to the $L = 16, OA$ state yields suppression of Raman in the $L = 16 \text{ and } 17$ OA modes as before (light and dark green curves in Fig. S5B). Finally, Stokes light in the $L = 15$ mode yields a similarly modulated Raman spectrum as that shown in Fig. S5A, with the additional presence of FWM peaks in its spectral proximity. This proves that the large error bars for selected data points of Fig. 3A arising from FWM do not otherwise affect our central conclusions, since the influence of FWM is clearly distinct and separable from that of the Raman scattering process we have aimed to study.

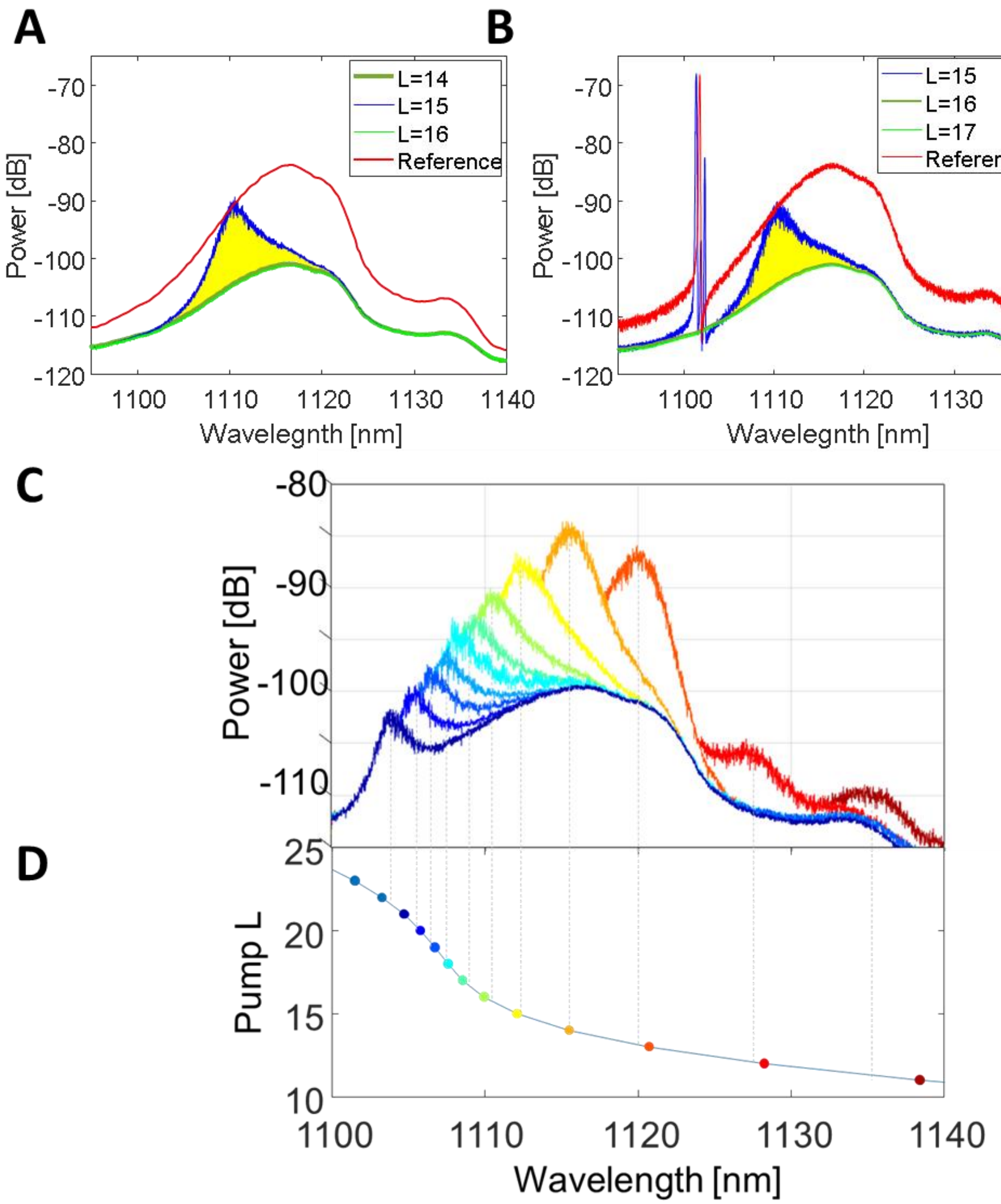

**Fig. S5. (A)** Simulated spectra based on the multi-mode nonlinear Schrödinger equation, with FWM intentionally excluded by choice of modes. Red curve corresponds to pump in $L = 16, \hat{\sigma}^{+}$, Stokes in $L = 15$, dark and light green curves correspond to pump in $L = 16$, OA, Stokes in $L = 14$ and 16, blue curve corresponds to $L = 16$, OA, Stokes in $L = 15$, the yellow shaded regime indicates the tailored Raman spectral shape. **(B)** Comparison with the case where FWM is included, red curve corresponds to pump in $L = 16, \hat{\sigma}^{+}$, Stokes in $L = 15$, dark and light green curves correspond to pump in $L = 16$, OA, Stokes in $L = 16$ and 17, blue curve corresponds to $L = 16$, OA, Stokes in $L = 15$. **(C)** Raman spectra in the beat length matching mode for pump in different $L$’s. **(D)** Simulated beat length matching wavelength as a function of pump $L$, black dashed line indicates good agreement with the modulated Raman gain peak wavelengths.